\documentclass[prd,amsmath,amssymb,groupedaddress,letterpaper,twocolumn]{revtex4}
\usepackage{amsmath,amssymb,graphicx,bm} 

\newcommand{\cH}{{\cal H}}

\newcommand{\cK}{{\cal K}}
\newcommand{\cL}{{\cal L}}

\newcommand{\cV}{{\cal V}}
\renewcommand{\d}{\delta}

\newcommand{\nn}{\nonumber}
\newcommand{\p}{\partial}

\begin{document}

\title{Instabilities and the null energy condition}
 
\author{Roman~V.~Buniy}
\email{roman@uoregon.edu}
\author{Stephen~D.H.~Hsu}
\email{hsu@duende.uoregon.edu}

\affiliation{Institute of Theoretical Science, University of Oregon,
  Eugene OR 94703-5203}

\begin{abstract}
We show that violation of the null energy condition implies
instability in a broad class of models, including classical gauge
theories with scalar and fermionic matter as well as any perfect
fluid. When applied to the dark energy, our results imply that $w = p
/ \rho$ is unlikely to be less than $-1$.
\end{abstract}
 
\maketitle

\section{Introduction}\label{I}

Energy conditions, or restrictions on the matter energy-momentum
tensor $T_{\mu \nu}$, play an important role in general relativity. No
classification of the solutions to Einstein's equation is possible
without restrictions on $T_{\mu \nu}$, since every spacetime is a
solution for some particular choice of energy-momentum tensor.  In
this letter we demonstrate a direct connection between stability and
the null energy condition (NEC) \cite{null}, $T_{\mu \nu} n^\mu n^\nu
\ge 0$ for any null vector $n$ (satisfying $g_{\mu\nu}n^\mu n^\nu =
0$). Our main results are: (1) classical solutions of scalar-gauge
models which violate the NEC are unstable, (2) a quantum state
(including fermions) in which the expectation of the energy-momentum
tensor violates the NEC cannot be the ground state, (3) perfect fluids
which violate the NEC are unstable. These results suggest that violations of
the NEC in physically interesting cases are likely to be only ephemeral.

Our results have immediate applications to the dark energy equation of
state, often given in terms of $w=p/\rho$. Dark energy has positive
energy density $\rho$ and energy-momentum tensor $T_{\mu \nu} = \text
{diag\,}(\rho, p, p, p)$ in the comoving cosmological
frame. Therefore, $w < -1$ implies violation of the NEC. Instability
as a consequence of $w < -1$ was studied previously in scalar
models~\cite{instability}.

Some results in relativity in which the NEC plays an important role
include the classical singularity theorems \cite{Hawking-Ellis},
proposed covariant entropy bounds \cite{bousso} and non-existence of
Lorentzian wormholes~\cite{wormholes}.

\section{Field theories}\label{FT}

Consider a theory of scalar, $\phi_a$, and gauge, $A_{a\alpha}$,
fields in a fixed $d$-dimensional space-time with the metric
$g_{\mu\nu}$. We limit ourselves to theories whose equations of motion
are second order differential equations, so the Lagrangian for the
system is assumed to depend only on the fields and their first
derivatives. We take the Lagrangian density $\cL$ to depend only on
the covariant derivative of the field $D_\mu\phi_a$ and the gauge
field strength $F_{a\mu\nu}$. The scalars may transform in any
representation of the gauge group. We impose Lorentz invariance on
$\cL$, but do not require overall gauge invariance. That is, we allow
for fixed tensors with gauge indices (but no Lorentz indices) which
can be contracted with the fields. For the corresponding action
\begin{eqnarray}
  S=\int
  d^dx\,|g|^\frac{1}{2}\cL(\phi_a,D_\mu\phi_a,F_{a\mu\nu})\label{FT:S}
\end{eqnarray}
to be stationary, its first variation has to vanish, $\delta
S=0$. This leads to the equations of motion for the fields $\phi_a$ and
$A_{a\alpha}$; in the classical analysis we assume that we have found
solutions to these equations, about which we expand.

\subsection{Null energy condition}\label{S:NEC}

The quantities $D_\mu\phi_a$ and $g^{\mu\nu}$ are independent
variables. Nevertheless we now prove that there is a relation between the
derivatives of $\cL$ with respect to them:
\begin{eqnarray}
  2\cL_{g^{\mu\nu}} &=& M^{AB}\psi_{A\mu}\psi_{B\nu} + g_{\mu\nu} K,
  \label{FT:M.g}\\ \cL_{\psi_{A\mu}} &=& M^{AB}{\psi_B}^\mu +
  \epsilon^{\mu \nu_2 \ldots \nu_{d}} {L^A}_{\nu_2 \ldots \nu_{d}}.
  \label{FT:M.psi}
\end{eqnarray}
In our notation $\psi_{A\mu}=(D_\mu\phi_a,F_{a\alpha\mu})$, where the
abstract index $A$ may run over both Lorentz and color indices, as
well as the type of field. So, $\psi_{A\mu}$ is a list of objects,
each of which has a Lorentz index $\mu$. The value of $A$ specifies an
element of this list.

The relations are obtained
by noting that for each and every $g^{\mu\nu}$ in $\cL$ there are two
$\psi$s attached to it, except for the curved space totally
antisymmetric tensor $\vert g \vert^{-\frac{1}{2}} \epsilon^{\nu_1
 \ldots \nu_d}$, which gives rise to the $K$ term in
Eq.~(\ref{FT:M.g}).  Similarly, differentiation with respect to
$\psi_{A\mu}$ yields the $M$ and $L$ terms in
Eq.~(\ref{FT:M.psi}). 

\begin{figure}[h!]
\includegraphics[width=6cm]{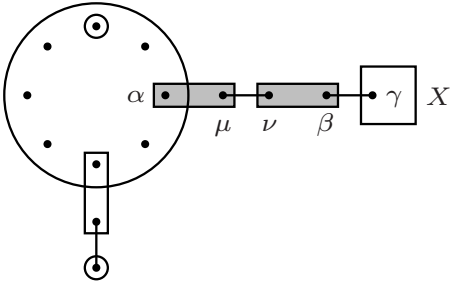}
\caption{Representation of the most general Lagrangian of the type
considered in this paper. Each dot represents a Lorentz index, and a
line connecting them denotes contraction using the metric. Rectangles
(with two indices) are field strengths, small circles covariant
derivatives of scalar fields, and large circles epsilon
tensors. Finally, the block $X$ represents the remainder of the
diagram. All Lorentz indices are ultimately contracted, and we
suppress color indices for simplicity.  In graphical terms, $M$ is
obtained by simply removing the shaded elements.}
\label{figure}
\end{figure}

Figure~\ref{figure} represents the most general Lagrangian of type
considered in this paper. Each dot represents a Lorentz index and a
line connecting them denotes contraction using the metric. Rectangles
(with two indices) are field strengths, small circles covariant
derivatives of scalar fields, and a large circle an epsilon
tensor. Finally, the block $X$ represents the remainder of the
diagram. Because the product of two epsilon tensors can be rewritten
as a sum of products of metric tensors $g$, we need to consider only
figures with at most one epsilon tensor, and therefore can assume that
$X$ contains none. (For generality, we include an epsilon tensor in
the figure, although of course $\cL$ need not contain a
parity-violating component.) All Lorentz indices are ultimately
contracted, and we suppress color indices for simplicity.

Consider the labelled portion of the figure, which equals
\begin{equation}
\cL = \vert g \vert^{-\frac{1}{2}} \epsilon^{\alpha \ldots}
F_{\alpha \mu} \, g^{\mu \nu} F_{\nu \beta} \, g^{\beta \gamma}
X_{\gamma \ldots}.
\label{FT:M.L}
\end{equation}
Again, we suppress color indices for simplicity, as they do not affect
the proof. We include an epsilon tensor in the analysis, although
$\cL$ may or may not contain one (in the parity-preserving case the
epsilon tensor in Eq.~(\ref{FT:M.L}) is replaced by the metric). By
differentiation of the indicated portion, we obtain
Eqs.~(\ref{FT:M.g}) and (\ref{FT:M.psi}) with
\begin{eqnarray}
M^{\alpha \beta} &=& - \vert g \vert^{-\frac{1}{2}} \left(
\epsilon^{\alpha \ldots} g^{\beta \gamma} + \epsilon^{\beta \ldots}
g^{\alpha \gamma} \right) X_{\gamma \ldots}, \\ K &=& \vert g
\vert^{-\frac{1}{2}} \epsilon^{\alpha \ldots} F_{\alpha \rho} \,
g^{\rho \sigma} F_{\sigma \beta} \, g^{\beta \gamma} X_{\gamma \ldots}, \\
{L^\alpha}_{\ldots} &=& - \vert g \vert^{-\frac{1}{2}} g^{\alpha \rho}
F_{\rho \beta} \,g^{\beta \gamma} X_{\gamma \ldots}. \label{FT:L}
\end{eqnarray}
Note the derivative $\cL_{F_{\nu \beta}}$ generates a contribution to
$M$ which is matched by a corresponding contribution from
$2 \cL_{g^{\beta \gamma}}$. Other contractions of fields with $g_{\mu
\nu}$ (i.e., as indicated in the figure) can be analyzed similarly.
In graphical terms, $M$ can be obtained from the figure for $\cL$ by simply
removing two $\psi$s, in this case the shaded elements of the figure.

For the energy-momentum tensor which couples to gravity,
\begin{eqnarray}
  T_{\mu\nu} =-\cL g_{\mu\nu} +2\cL_{g^{\mu\nu}}, \label{FT:T}
\end{eqnarray}
the NEC then requires
\begin{eqnarray}
  \Psi_A M^{AB}\Psi_B\ge 0, \label{FT:NEC.M}
\end{eqnarray}
where $\Psi_A=\psi_{A\mu}n^\mu$. Thus, to satisfy the NEC, $M^{AB}$ has to
be positive semidefinite. This property is crucial for stability of
solutions, to which we now turn.

\subsection{Stability}\label{S:Stability}

To study the stability of the solution $\psi_A(x)$, we consider the
second variation of the Lagrangian,
\begin{eqnarray}
  \d^2 \cL &=& \cL_{\psi_A\psi_B}
  \delta\psi_A\delta\psi_B  + 
  2\cL_{\psi_{A}\psi_{B;\lambda}}\delta\psi_{A}\delta\psi_{B;\lambda}
  \nn \\ &+& \cL_{\psi_{A;\mu}\psi_{B;\nu}}
  \delta\psi_{A;\mu}\delta\psi_{B;\nu}.\label{FT:d2L}
\end{eqnarray}
Here quantities $\cL_{\psi_A}=\p\cL/\p\psi_A$, etc. are evaluated at
$\psi_A(x)$. Also notice that
$\psi_{A;\mu}=(D_\mu\phi_a,A_{a\alpha;\mu})$, the covariant
derivatives of $\psi_A$, are different from
$\psi_{A\mu}=(D_\mu\phi_a,F_{a\alpha\mu})$.

Let us use a locally inertial frame in which the metric is reduced to
$\bar{g}_{\mu\nu} =\text{diag\,}(1,-1,\ldots,-1)$; all quantities in
this frame are designated by a bar. For the
Lagrangian~(\ref{FT:d2L}), the canonical momentum is
\begin{eqnarray}
\delta\bar{\pi}^A =
2\cL_{\bar\psi_{B}\bar\psi_{A;0}}\delta\bar\psi_{B} + 
2 \cL_{\bar\psi_{A;0}\bar\psi_{B;\nu}} \delta\bar\psi_{B;\nu}
\label{FT:dpi}
\end{eqnarray}
which leads to the following effective Hamiltonian for fluctuations
about the classical solution:
\begin{eqnarray}
\delta^2\cH &=& -\cL_{\bar\psi_{A}\bar\psi_{B}} \delta\bar\psi_{A}
\delta\bar\psi_{B} -2\cL_{\bar\psi_{A}\bar\psi_{B;j}}
\delta\bar\psi_{A} \delta\bar\psi_{B;j} \label{FT:d2H} \\ &+&
\cL_{\bar\psi_{A;0}\bar\psi_{B;0}} \delta\bar\psi_{A;0}
\delta\bar\psi_{B;0} - \cL_{\bar\psi_{A;i}\bar\psi_{B;j}}
\delta\bar\psi_{A;i} \delta\bar\psi_{B;j}.\nn
\end{eqnarray}
Here $\delta\bar{\psi}_{A;0}$ are functions of $\delta\bar{\pi}^B$,
$\delta\bar{\psi}_{B}$ and $\delta\bar{\psi}_{B;i}$ as found from
Eq.~(\ref{FT:dpi}). The first term on the right hand side of
Eq.~(\ref{FT:d2H}) is a potential term, which we denote by
$\delta^2\cV$, and the last two terms are kinetic terms, denoted
$\delta^2\cK$.

If the kinetic energy $\delta^2\cK$ is
negative, then the system described by the Hamiltonian $\delta^2\cH
=\delta^2\cK +\delta^2\cV$ is (locally) unstable. 
If $\delta^2\cV$ is positive then small
perturbations will cause the classical solutions to grow
exponentially away from the original stationary point. However, it is
possible to have classical stability if one chooses $\delta^2\cV$ to
be negative; in this case we have an upside-down potential with
negative kinetic term, or a ``phantom''. Such models necessarily
exhibit quantum instabilities \cite{phantom}.
Notice, the second term in Eq.~(\ref{FT:d2H}) is linear in the
fluctuations and their derivatives, and therefore can never
stabilize the system.

To investigate the kinetic terms, we calculate second derivatives of
$\cL$ needed in Eq.~(\ref{FT:d2H}). Using Eq.~(\ref{FT:M.psi}) we
obtain
\begin{equation}
\cL_{\psi_{A\mu}\psi_{B\nu}} = M^{AB}g^{\mu\nu} +N^{A\mu
B\nu}.\label{FT:MN}
\end{equation}
The separation of the second derivative into the first and second
terms in Eq.~(\ref{FT:MN}) is natural: $g^{\mu\nu}$ appears only in
the first term, and $N$ represents all remaining terms. $N$ contains
terms obtained by differentiating $M$ and $L$ with respect to
$\psi_{B\nu}$, plus additional terms if $\psi$ is a field
strength. The $\nu$ index obtained from these $\psi_{B\nu}$
derivatives is attached to a field and not the metric $g^{\mu
\nu}$. (Also, $L$ does not contain an epsilon tensor since $X$ does not.)
Finally, notice that even though $\psi_{A\mu}$ and $\psi_{A;\mu}$ differ,
the derivatives of $\cL$ with respect to them coincide due to the form
of the action~(\ref{FT:S}). Thus the kinetic term becomes
\begin{eqnarray}
\delta^2\cK &=& \left(\bar{M}^{AB}+\bar{N}^{A0B0}\right)
    \delta\bar\psi_{A;0}\delta\bar\psi_{B;0}\nn\\ &+&
    \left(\bar{M}^{AB}\delta^{ij} -\bar{N}^{AiBj}\right)
    \delta\bar\psi_{A;i}\delta\bar\psi_{B;j}.\label{FT:d2K}
\end{eqnarray}

We now prove that nonnegativeness of the kinetic term $\delta^2\cK$
implies positive semidefiniteness of the matrix $M$. Indeed,
suppose that $M$ is not positive semidefinite. In such case, the
matrix $M$ has at least one negative eigenvalue, which means that
there is a basis in which the matrix is diagonal with at least one
negative entry,
$\tilde{M}=\text{diag\,}(\tilde{m}_1,\ldots,\tilde{m}_n)$
($\tilde{m}_1<0$). (Quantities in this basis are designated with a
tilde.) Let us choose such field variations that are nonzero
only in the direction of the negative eigenvalue:
$\delta\tilde{\psi}_{1;\mu}\not =0$ and $\delta\tilde{\psi}_{A;\mu}=0$
$(A>1)$. We further restrict $d-1$ quantities
$\delta\tilde{\psi}_{1;i}$ to satisfy the following equation:
\begin{eqnarray}
  \tilde{N}^{1010}\delta\tilde{\psi}_{1;0} \delta\tilde{\psi}_{1;0}
  =\tilde{N}^{1i1j}\delta\tilde{\psi}_{1;i}\delta\tilde{\psi}_{1;j}.
  \label{FT:Ncondition}
\end{eqnarray}
These conditions make the kinetic term of Eq.~(\ref{FT:d2K}) negative,
\begin{eqnarray}
  \delta^2\cK =\tilde{m}_1\left[(\delta\tilde{\psi}_{1;0})^2
    +\sum_i(\delta\tilde{\psi}_{1;i})^2\right] <0,
  \label{FT:d2K1}
\end{eqnarray}
thus proving that in order for $\delta^2\cK$ to be nonnegative, the
matrix $M$ has to be positive semidefinite.
 
Using the result established in the previous paragraph, we conclude
that solutions to the theory given by the action~(\ref{FT:S}) are
stable only if the matrix $M$ is positive semidefinite.
 
Combining the relations between nonnegativeness of $\delta^2\cK$ and
positive semidefiniteness of $M$ on one hand, and the NEC and
positive semidefiniteness of $M$ on the the other hand, we
conclude that for the theory given by the action~(\ref{FT:S}), only
solutions satisfying the NEC can be stable.

We can deduce similar results for quantum systems. Suppose there
exists a quantum state $\vert \alpha \rangle$ and a null vector $n^\mu$
such that
\begin{equation}
\langle \alpha \vert T_{\mu\nu} \vert \alpha \rangle n^{\mu} n^{\nu} =
\langle \alpha \vert M^{AB} \Psi_{A} \Psi_{B} \vert \alpha \rangle <
0,
\end{equation}
so that the NEC is violated in a quantum averaged sense.  Define a
basis $\vert \phi \rangle$ in which the operator ${\cal M} = M^{AB}
\Psi_A \Psi_B$ is diagonal:
\begin{equation}
{\cal M} \vert \phi \rangle = m(\phi) \vert \phi \rangle.
\end{equation}
Then violation of the quantum averaged NEC implies
\begin{equation}
\sum_{\phi \phi'} \, \langle \alpha \vert \phi \rangle \langle \phi
\vert {\cal M} \vert \phi' \rangle \langle \phi' \vert \alpha \rangle
\\ = \sum_{\phi} \, \vert \langle \alpha \vert \phi \rangle \vert^2
\,m(\phi) < 0.
\end{equation}
This means that there exist eigenstates $\vert \phi \rangle$, whose
overlap with $| \alpha \rangle$ is non-zero and on which the operator
${\cal M}$ has negative eigenvalues. This requires that $M$ and
hence $\delta^2\cK$ is not positive semidefinite; by continuity, this
must also be the case in a ball $B$ in the Hilbert space of $\vert \phi
\rangle$.

As a further consequence, we can conclude that a state $\vert \alpha
\rangle$ in which the NEC is violated cannot be the ground
state~\footnote{In some curved spacetimes there may not be a well-defined
ground state. In de Sitter space, quantum corrections can cause
violation of the NEC: V.~K.~Onemli and R.~P.~Woodard, Class.\ Quant.\
Grav. {\bf 19}, 4607 (2002); Phys.\ Rev.\ D {\bf 70}, 107301
(2004).}. Suppose that $\vert \alpha \rangle$ is an energy eigenstate:
$H \vert \alpha \rangle = E_\alpha \vert \alpha \rangle$. An
elementary result from quantum mechanics is that $\vert \alpha
\rangle$ can be the ground state only if
\begin{equation}
E_\alpha = \langle \alpha \vert H \vert \alpha \rangle \leq \langle
\alpha' \vert H \vert \alpha' \rangle
\end{equation} 
for all normalized states $\vert \alpha' \rangle$ which need not be
energy eigenstates. However, it is possible to reduce the expectation
value of $H$ by perturbing $\vert \alpha \rangle$. Specifically, we
adjust $\vert \alpha \rangle$ only in the ball $B$, where we know from
Eqs.~(\ref{FT:d2K})--(\ref{FT:d2K1}) that there are perturbations
which reduce the expectation of the kinetic energy without changing
the expectation of the potential. 

Note that the discussion above is in terms of unrenormalized (bare)
quantities. The renormalized expectation $\langle \alpha \vert {\cal
M}_{\rm ren} \vert \alpha \rangle = \langle \alpha \vert {\cal M}
\vert \alpha \rangle - \langle 0 \vert {\cal M} \vert 0 \rangle$
(where $\vert 0 \rangle$ is the flat-space QFT ground state) could be
negative (e.g., as in the Casimir effect \cite{null}), but this is
possible only if $\vert \alpha \rangle$ is not $\vert 0 \rangle$.

In known cases of NEC violation, such as the Casimir vacuum or black
hole spacetime, it is only the {\it renormalized} energy-momentum
tensor which violates the NEC. As a simple example, consider a real
scalar field $\phi$. The energy-momentum tensor is simply $T_{\mu \nu}
= \partial_\mu \phi \, \partial_\nu \phi$ plus terms proportional to
$g_{\mu\nu}$ which do not play a role in the NEC. Then, ${\cal M} = (
n^{\mu} \partial_{\mu} \phi )^2$ is a Hermitian operator with positive
eigenvalues. Therefore, its expectation value in {\it any} state is
positive: $\langle \alpha \vert {\cal M} \vert \alpha \rangle > 0,$
for any $\vert \alpha \rangle$, including the Hartle-Hawking, Casimir
or flat-space vacuum. We can verify this by direct calculation,
computing the energy-momentum tensor using point-splitting
regularization:
\begin{eqnarray}
\langle 0 \vert T_{\mu\nu} (x,x') \vert 0 \rangle n^\mu n^\nu = \frac{2
[n^\mu (x-x')_\mu]^2 }{\pi^2 \vert x-x'\vert^6},
\end{eqnarray}
which is manifestly positive. Note that $\langle \alpha \vert {\cal M}
\vert \alpha \rangle > 0$ for all $\vert \alpha \rangle$, since the
bare expectation is always dominated by the UV contribution. Now, had
we taken a {\it negative} kinetic energy term for the scalar, the
overall sign of ${\cal M}$ would change, allowing violation of the
NEC. But, this model is clearly unstable, in accordance with our
results.

\subsection{Fermions}\label{FF}

To this point we have only considered bosonic fields. We now extend
our analysis to systems with fermions, adding to our Lagrangian the
term
\begin{equation}
\cL^{(\text{f})} = \bar{\psi} ( i D\!\!\!\!/ - m ) \psi.
\end{equation}
(A scalar-fermion coupling can be treated similarly, as can a Weyl
fermion, whose determinant is the square root of the Dirac
determinant.) Then, for any fixed gauge field background the fermions
can be integrated out directly in favor of a non-local correction to
the action for bosonic fields
\begin{equation}
- \sum_{\lambda_l > 0} \ln ( \lambda_l^2 + m^2 ),
\end{equation}
where $D\!\!\!\!/ \, \psi_l = \lambda_l \psi_l$ is the eigenvalue
equation for the Dirac operator, with $\lambda_l$ real. This shifts
the energy-momentum tensor by

\begin{equation}
  T_{\mu \nu}^{(\text{f})} = -2\vert g \vert ^{-\frac{1}{2}}
    \sum_{\lambda_l > 0} \frac{1}{\lambda_l^2 + m^2} \frac{\d
    \lambda_l^2}{\d g^{\mu \nu}}.
\end{equation}
Now write $\lambda_l^2 \psi_l = {D\!\!\!\!/ }^{\,2} \psi_l = (g^{\mu
\nu} D_\mu D_\nu -\tfrac{1}{2}i \sigma^{\mu\nu} F_{\mu\nu} )\psi_l $
and use the orthonormality of the eigenfunctions to obtain
\begin{equation}
\lambda_l^2 = \int d^dx \, |g|^\frac{1}{2} \, g^{\mu \nu}
\,\psi^\dagger_l D_\mu D_\nu \psi_l ~+~ \ldots,
\end{equation}
where the ellipsis denote terms which do not contain $g^{\mu\nu}$.
After integration by parts, the contribution to the NEC from fermions
is then
\begin{equation}
T^{(\text{f})}_{\mu\nu} n^\mu n^\nu =  \sum_{\lambda_l >
0} \frac{2} {\lambda_l^2 + m^2} \, 
( n \cdot D \psi_l)^\dagger ( n \cdot D \psi_l ).
\end{equation}
This additional contribution is always positive. So, the conclusions
of the previous section are unmodified by the presence of fermions:
violation of the NEC implies the bosonic kinetic energy is not
positive semidefinite.

\section{Perfect fluid}\label{F}

A macroscopic system may be approximately described as a perfect fluid
if the mean free path of its components is small compared to the
length scale of interest. For the dark energy, this length scale is of
cosmological size.  A perfect fluid is described by the
energy-momentum tensor
\begin{eqnarray}
T_{\mu\nu}=(\rho+p)u_\mu u_\nu-pg_{\mu\nu},\label{F:T}
\end{eqnarray}
where $\rho$ and $p$ are the energy density and pressure of the fluid
in its rest frame, and $u_\mu$ is its velocity. Let $j^\mu=J u^\mu$ be
the conserved current vector (${j^\mu}_{;\mu}=0$), and $J=(j_\mu
j^\mu)^\frac{1}{2}$ the particle density.

The energy-momentum can be written \cite{Hawking-Ellis,Jackiw:2004nm}
as
\begin{eqnarray}
T_{\mu\nu}=(f-J f')g_{\mu\nu}+(f'/J) j_\mu j_\nu,\label{F:T1}
\end{eqnarray}
where, comparing with Eq.~(\ref{F:T}), we have $\rho = f(J)$ and $p =
Jf' - f$. The function $f(J)$ implicitly determines the equation of state.  
 
The NEC for the tensor~(\ref{F:T1}) becomes
\begin{eqnarray}
T_{\mu\nu}n^\mu n^\nu=(f'/J)(j_\mu n^\mu)^2\ge 0.\label{F:NEC}
\end{eqnarray}
Thus, perfect fluids with negative $f'(J)$ violate the NEC. Below, we
demonstrate that $f'(J) < 0$ implies an instability.

Recall that the speed of sound in a fluid is given by $s = (dp /
d\rho)^{\frac{1}{2}} = (J f'' / f')^{\frac{1}{2}}$, and that complex
$s$ implies an instability. Note $f'(J)$ cannot change its sign
without producing an instability. Indeed, if it were to change its
sign at some $J_*$, then $s$ would be complex for either $J$ larger or
smaller than $J_*$, depending on the sign of $f''(J_*)$.  ($f''$
cannot also change sign at $J_*$.)  Therefore, if $f'$ is negative
anywhere, then it is negative everywhere, and to avoid complex $s$,
$f''$ must be negative everywhere.

However, if $f'$ and $f''$ are everywhere negative, then the fluid is
unstable with respect to clumping. To see this, we first deduce the
dependence of the fluid free energy $F$ on particle number $N =
JV$. Note that $( \partial F / \partial V )\vert_{T,N} = - p = N
\partial (f/J) / \partial V$. By integration we find $F = N [ f/J -
h(T)]$, where the first term is just the energy $E$ and $N h(T) = T
S$, where $S$ is the entropy. It is easy to see that $(\partial^2 F
/\partial J^2 )\vert_{T,V} = V f''$.

Now consider two adjacent regions of the fluid
with identical volumes. Suppose we transfer a small amount of matter
$\delta J$ from one volume to another; the resulting change in total
free energy is given by $\tfrac{1}{2} V f''(J) (\delta J)^2 <
0$. We see that the system can decrease its free energy by clumping into
over- and under-dense regions. This itself is an instability, which
results in a runaway to infinitely negative free energy unless the
assumption of negative $f'$ (violation of NEC) or negative $f''$ (real
$s$) ceases to hold.

\bigskip

\begin{center}
\textbf{Acknowledgments}
\end{center}

The authors thank R.~Bousso, A.~Jenkins, B.~Murray, M.~Schwartz,
D.~Soper and M.~Wise for useful comments. This work was supported by
the Department of Energy under DE-FG06-85ER40224.

\end{document}